# In-situ tuned photoelectric properties of PtS$_2$ transistor

Yana Cui[1], Wentao Gong[1], Gang Zhao[1] and Weike Wang[1a)]

[1] *Synergetic Innovation Center for Quantum Effects and Application, Key Laboratory of Low-dimensional Quantum Structures and Quantum Control of Ministry of Education, College of Physics and Electronics Science, Hunan Normal University, Changsha, 410081, Hunan, China*

a) To whom correspondence should be addressed. E-mail: wkwang@hunnu.edu.cn or yanacui@hunnu.edu.cn

# Abstract

Strain engineering is a powerful and widely used strategy for boosting the performance of electronic and optoelectronic devices. Here, we demonstrate an approach to tune the photoelectric properties of Platinum sulfide ($PtS_2$) by using a ferroelectric substrate PMN-PT as the strain generator. It is found that both the drain current and responsivity of the $PtS_2$ photodetector is directly coupled to the electrostriction of PMN-PT, showing a high strain-tuned ratio $\sim 10^3$, high responsivity up to $6.3 \times 10^3$ A/W and detectivity of $9.3 \times 10^{12}$ Jones. Additionally, a high photogain $\approx 5 \times 10^5$ is obtained at a gate voltage $V_g = 15$ V. Our results provide an effective method for manipulating electrical properties and optimizing performance of two dimensional layered (2D) materials based optoelectronic devices.

**Introduction**

The layered two-dimensional material has attracted more and more attention in recent years because of its physical properties and promising potential applications in the field of field-effect transistors, optoelectronics devices, thermoelectric devices and so on[1-4]. Among them, many 2D semiconductor photodetectors have been demonstrated with excellent performance. For example, $Ta_2NiSe_5$ photodetector attains ultrahigh photoresponsivity up to 17.21 A / W[5], and GeP phototransistor achieves highly anisotropic photoresponsivity[6]. It is well known that the transistor performance was largely depredated due to scattering from charged surface states and impurities, surface roughness and so on. Thus, it improves the performance of transistors through external physical methods, for instance, by decorating Au nanoparticles on the surface, $MoS_2$ phototransistor exhibits a threefold enhancement in the photocurrent[7]. The $PtS_2$ phototransistor on h-BN substrates shows a sharply increasing responsivity[8]. Usually strain can effectively tune the physical properties of the material, by applying a strain through lattice mismatch between epitaxial films and substrates or through bending of films on elastic substrates. Applying strain can be used to constructe an in-situ controlled photodetector, which will benefit to designing and optimizing the performance of 2D photodetectors in practical applications.

Recently, as one of the tenth transition metal dichalcogenide, $PtS_2$, has attracted the attention of many scientists due to its high mobility and air stability[10-14]. Interestingly, $PtS_2$ shows a layer-dependent band gap from 1.6 eV of monolayer to 0.25 eV of bulk[15]. Moreover, theoretical calculations have demonstrated that the band gap of $PtS_2$ can be finely modulated under mechanical strain[16,17]. In experimental, Yuan et al. explored photoelectric performance of $PtS_2$ under pressure study, showing the pressure-enhanced photocurrent[18]. Those research results indicate the possibility of modulating photoelectric properties of $PtS_2$ by means of applying strains. The relaxation type ferroelectric single crystal $0.72PbMg_{1/3}Nb_{2/3}O_3$-$0.28PbTiO_3$ (PMN-PT) has a high piezoelectric constant and relatively high piezoelectric activity[19,20], and can serve as strain generator controlled by the external electric field[21]. In this work, we employed piezoelectric materials PMN-PT as the strain generator. By mounting an PtS$_2$ ultrathin flake onto the PMN-PT substrate, we demonstrated that the drain current is directly tuned by the electrostriction of PMN-PT. The in-situ strains are fined manipulated by an external electrical field. We systematically studied the photoelectric performance of the device on PMN-PT substrate. The responsivity can be tuned up to 6.3 × 10$^3$ A / W. Using strain-engineered 2D materials with ferroelectrics represents a fundamentally exciting platform to explore the wide variety of other electric-field-tuned photoelectronic properties in the 2D materials world.

In order to study the strain effect on electron transport properties of PtS$_2$, we exfoliated bulk PtS$_2$ to obtain ultrathin flakes by scotch-tape based micromechanical exfoliation method, and transferred onto monocrystal PMN-PT substrate with (001) orientation and the thickness of 0.5 mm. The PtS$_2$ device was fabricated by standard electron beam lithography (EBL), and Ti / Au (10 nm / 80 nm) was deposited as the contact electrode. The schematic diagram of the device is illustrated in figure 1(a). The PtS$_2$ device consists of four electrodes, in which two electrodes were not contacted to the sample acted as Gate Electrodes, and the others were the source-drain electrodes. Figure 1 (b) shows an atomic force microscope image of a typical PtS$_2$ device. The thickness of the sample is about 36 nm in figure 1 (c). The Raman spectrums of PtS$_2$ measured on PMN-PT and on SiO$_2$ / Si substrates as shown in figure 1 (d). The observed Raman peaks agrees with reported results[8], showing that the sample is intact and not damaged.

**Results and discussion**

Firstly, we performed electronic transport measurements in dark conditions. The electrical characterization of the PtS$_2$ transistor was performed using a semiconductor characterization system (4200SCS, Keithley) with a probe station (CRX-6.5K, Lake Shore). Figure 2 (a) displays the output characteristics $I_{ds}$-$V_{ds}$ of the PtS$_2$ device under different gate voltages ($V_g$). All $I_{ds}$-$V_{ds}$ characteristics show linear behavior in the region of approximately zero voltage, indicating that the Ti / Au electrode has good ohmic contact with the PtS$_2$ flake. In addition, the $I_{ds}$-$V_{ds}$ curve strongly depends on gate voltage ($V_g$), that is, the slope of the $I_{ds}$-$V_{ds}$ curve increases as $V_g$ increases, showing the strain-dependent conductance characteristics. We also measured the source-drain current as a function of the gate voltage measured at room temperature with $V_{ds}$ = 1 V, as shown in figure 2 (b). The $I_{ds}$ increases from 0.1 nA to 0.33 μA with increase of the gate voltage $V_g$, leading to a high strain-tuned ratio ~$10^3$. Note that the above device characteristic for $I_{ds}$-$V_g$ was obtained multiple times by cycling back and forth between "on" and "off" to ensure that the observed effects were reproducible and reversible. In our case, the gate voltage is applying across the PtS$_2$ channel, thus, we can reasonably ignore the ferroelectric polarization related change transfer. According to the piezoelectric constant of the substrate ($d_{33}$ ≈ 730 pm / V), it can be judged that the compressive stress of the sample is about 0.7 % with $V_g$ = 10 V[22]. The corresponding electric field induced a considerable strain due to electrostriction of PMN-PT can effectively pass on the PtS$_2$ flake owing to fine adsorption between flake and substrate. Thus, the band gap of the PtS$_2$ flake becomes narrow due to the PtS$_2$ flake compressed[23], and the carrier concentration in the conductive channel increases. The gate voltage increases, and the $I_{ds}$ rapidly raised. Those measurement results agree well with the result of theoretical prediction, clearly showing tunable electronic transport properties of PtS$_2$ by strain.

Then, we checked the optoelectronic properties of PtS$_2$ with remaining strain constant and all the measurements were carried out under five gate voltage. Figure 3 summarizes the optoelectronic properties of the PtS$_2$ device with $V_g$ = 5 V under λ = 445 nm laser illumination. As is shown in figure 3 (a)，the comparison of the output characteristics $I_{ds}$-$V_{ds}$ of PtS$_2$ under dark field and the laser irradiation clearly manifest the a net increase in the current. For further study, we measured the time-dependent photoresponse with different irradiation power intensities. Figure 3 (b) shows the *I-T* curves at different laser power intensities, one can clearly see that the $I_{ds}$ gradually increases with increasing power intensity. The photoresponsivity (*R*) is one of the most important figures of merit to evaluate the sensitivity of a photodetector. *R* can be determined as, $R=I_{ph}/PS$[24], where $I_{ph}$ is the photogenerated current, *P* is the power intensity and *S* is the effective irradiated area on the device. The calculated responsivity of the device is shown in figure 3 (c). From which, we can see that the responsivity achieves $4.45 \times 10^2$ A / W at a power intensity of 11.38 mW / cm$^2$ and decreases with increasing power intensity. The decrease in photoresponsivity at higher illumination intensities is due to trap states present inside PtS$_2$ or at the interface between PtS$_2$ and substrate[25,26]. Figure 3 (d) displays the cyclability of the PtS$_2$ device under power intensity of 21.83 mW / cm$^2$ by intentionally switching on / off the incident light, which clearly shows the stability of the on-off switching behavior. In the inset of figure 3 (d), one can see that the response time of the device is very short, that is, the rise time and fall time are 0.44 s and 0.42 s respectively. The switching speed is comparable or faster than many other reported 2D photoelectronic devices[27,28]. Noticed that those results are roughly in accordance with the previous reported[8], which is in favor of further study on strain effects.

For the further investigation of strain-tuned photoelectronic properties of the PtS$_2$ device, we performed systematical measurements on photocurrent as a function of power intensity and gate voltages. Figure 4 summarizes optoelectronic properties of PtS$_2$ device under electrical field induced strains. The relationships between the gate voltage $V_g$ and source-drains current $I_{ds}$ of the device at different illumination intensities are depicted in figure 4 (a). With increasing $V_g$ from 0 to 10 V, the substrate shrinks, corresponding to the compressed strains, and the $I_{ds}$ of the device sharply raised up, then approached a saturation current. However, upon $V_g$ decreasing from 0 to -5 V, the substrate expands, corresponding to the tensile strain, and the $I_{ds}$ rapidly reduced to the saturation value. The three-dimensional schematic diagram of the photocurrent is plotted in figure 4 (b). It is clearly seen that the evolution of the photocurrent with the gate voltage $V_g$ and the laser power, which exactly demonstrates the strain modulated optoelectrical properties of PtS$_2$. In our few-layered PtS$_2$ photodetector, changes in transport and photodetection behaviors with strain may arise from the piezoresistive effect, where strain results in changes in band gap structure and density of states of the carriers. When $V_g > 0$ V, a compressive strain is applied, the band gap becomes smaller[25]. With the decrease of the band gap, the absorption of photons with energy higher than the bandgap ($E_{ph} > E_{bg}$) can generate more free carriers, resulting in a net increase in the current.

Other figures of merit important to a photodetector, such as photoresponsivity, photogain, and detectivity, can also be modulated and largely enhanced by strain. We then study the photoresponsivity and detectivity of the device with strain, which is shown in figure 4 (c). It can be seen that the responsivity increases with increasing $V_g$, for $V_g$ = 15 V, the responsivity of the photodetector can reach up to 6.3 × 10$^3$ A / W, such value is two times higher than the previous reported[8]. Detectivity (D*) usually can be defined as D* = $R_\lambda S^{1/2}/(2eI_{dark})^{1/2}$, is another critical parameter to evaluate the detection performance of photodetectors. It can see that detectivity reaches 9.3 × 10$^{12}$ Jones at $V_g$ = 2 V for our device, which is 30 times higher than the previous reported (2.9 × 10$^{11}$ Jones)[8]. The photogain (G) is defined as the ratio between the number of electrons collected and the number of absorbed photons in the device per unit time. Figure 4d shows photogain of the device derived from data in figure 4b according to G = [$I_{ph}/(\eta P)$]($hv/q$)], where $\eta$ is the absorption of few-layered PtS$_2$, h is Planck's constant, q is the electronic charge, and v is the frequency of incident light. where η is the absorption of few-layered PtS$_2$, hv is the photon energy. Figure 4d shows that G increases with decreasing excitation laser power and reaches ≈ 5 × 10$^5$ when the excitation laser intensity is 14.83 mW / cm$^2$, if 4% absorption is estimated. These results suggest that strain is able to function as a controlling gate signal and effectively modulates the photodetection properties of few-layered PtS$_2$ optoelectronic device.

**Conclusion**

To summarize, we have successfully fabricated few-layered PtS$_2$ phototransistors on a PMN-PT substrate and studied photoelectric properties of PtS$_2$ under strain control. The responsivity of this phototransistor is 6.3 × 10$^3$ A / W, and its strain-tuned ratio is ~10$^3$. A maximum detectivity of 9.3 × 10$^{12}$ Jones with a 30-fold improvement over the reported highest photoresponsivity for few-layered PtS$_2$ phototransistors is demonstrated under strain. In addition, under compressive strain state at $V_g$ = 15 V, an ultrahigh photogain up to 5 × 10$^5$ is achieved. Therefore, PtS$_2$ will have unique advantages in the future photoelectric field due to its outstanding optoelectronic performance and the strain tuned photocurrent. Based on our results, incorporating strain engineering into future TMD devices occurs highly promising to tune the material properties of this class of two-dimensional materials.

**Acknowledgments**

This work was financially supported by the National Natural Science Foundation of China, Grant No.11604340, No. 11304321, and No. 11574081; and the project funded by China Postdoctoral Science Foundation, Grant No. 2015LH0018 and No. 2017M610474.

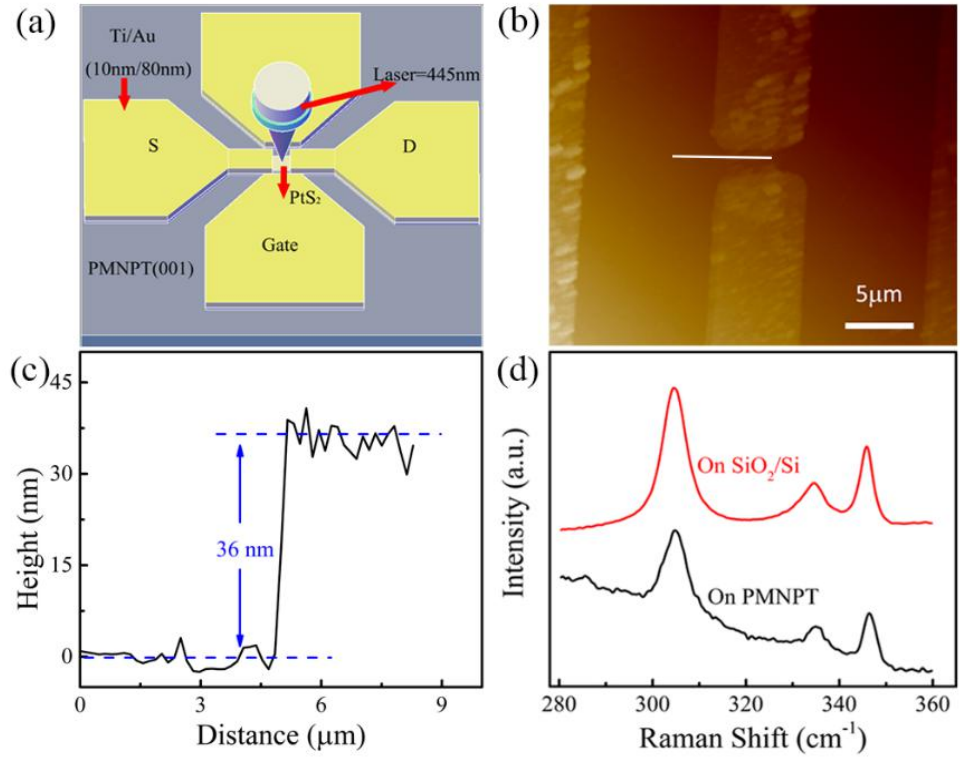

Figure 1. (a) Schematic structure of the fabricated PtS$_2$ device. (b) The atomic force microscope (AFM) image of a typical fabricated device. (c) The height profile along the white line of figure 1 (b) is shown here, and the thickness is about 36 nm. (d) Raman spectrum of PtS$_2$ on PMN-PT and on Si / SiO$_2$.

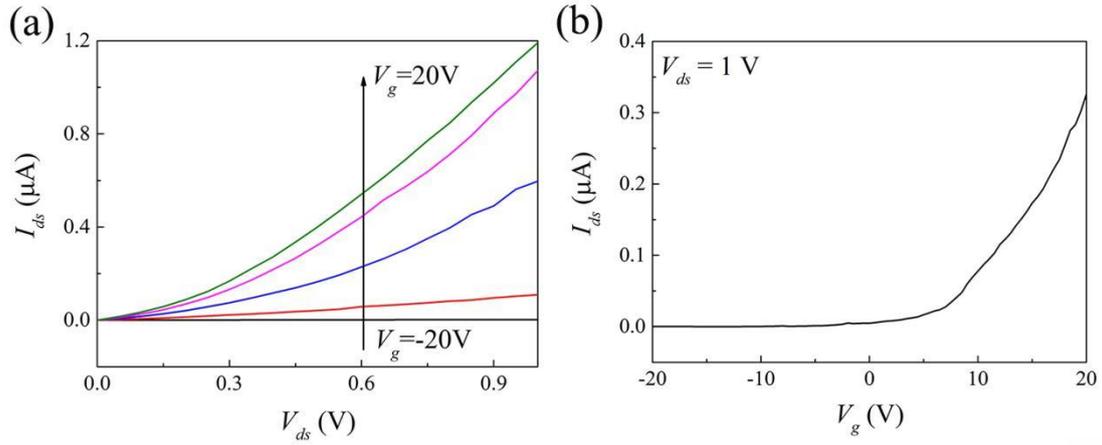

Figure 2. In the dark, the electronic transport properties of few-layered of PtS$_2$ based on PMN-PT. (a) Output characteristics $I_{ds}$-$V_{ds}$ of the device at different back-gate voltages. (b) Transfer curves $I_{ds}$-$V_g$ of the device with $V_g$ from -20 to 20 V at the drain-source bias $V_{ds}$ = 1 V.

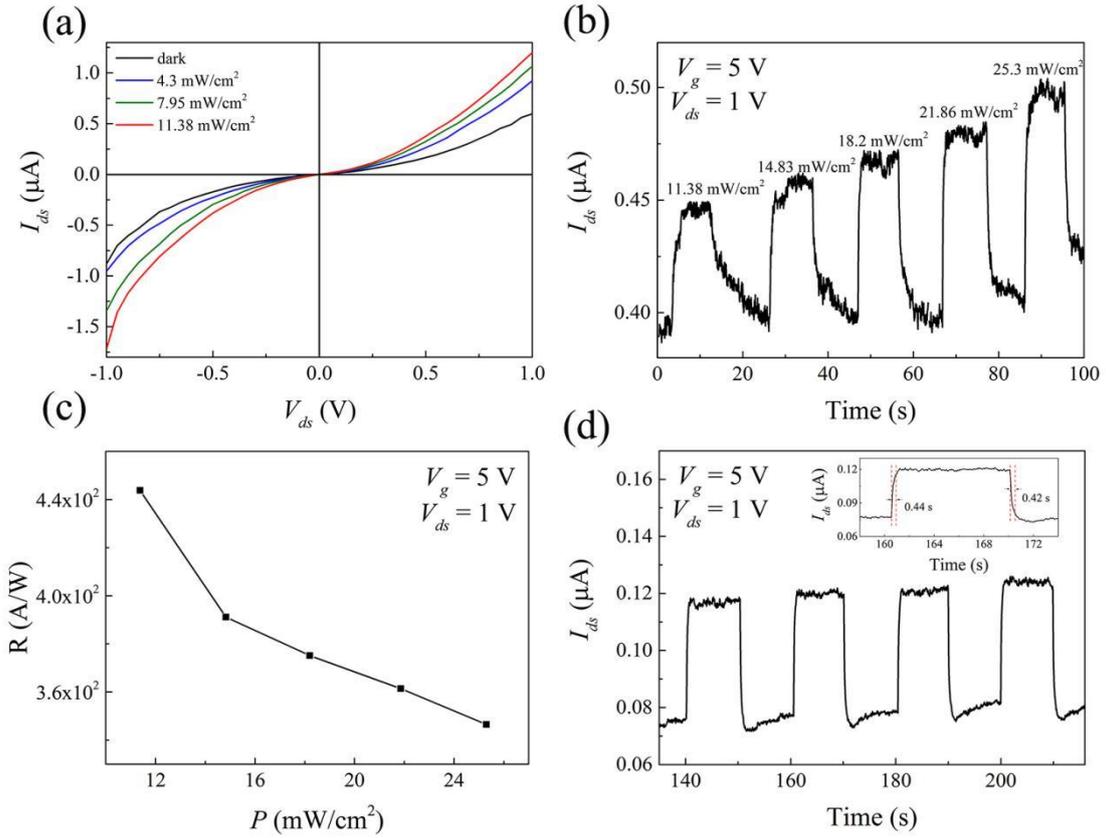

Figure 3. The photoelectric properties of few-layered of PtS$_2$ based on PMN-PT when $V_g$ = 5 V is applied (λ = 445 nm). (a) The output curves ($I_{ds}$-$V_{ds}$) of the phototransistor recorded in the dark and under different illumination intensities. (b) Light intensity-dependent photoresponse at $V_{bias}$ = 1 V. (c) The responsivity as a function of laser power intensity. (d) Timeresolved photoresponse recorded at $V_g$ = 5 V and $V_{ds}$ = 1 V, inset: response rate of photodetector acquired from one magnified circle of response with rising time 0.44 s and decay time 0.42 s.

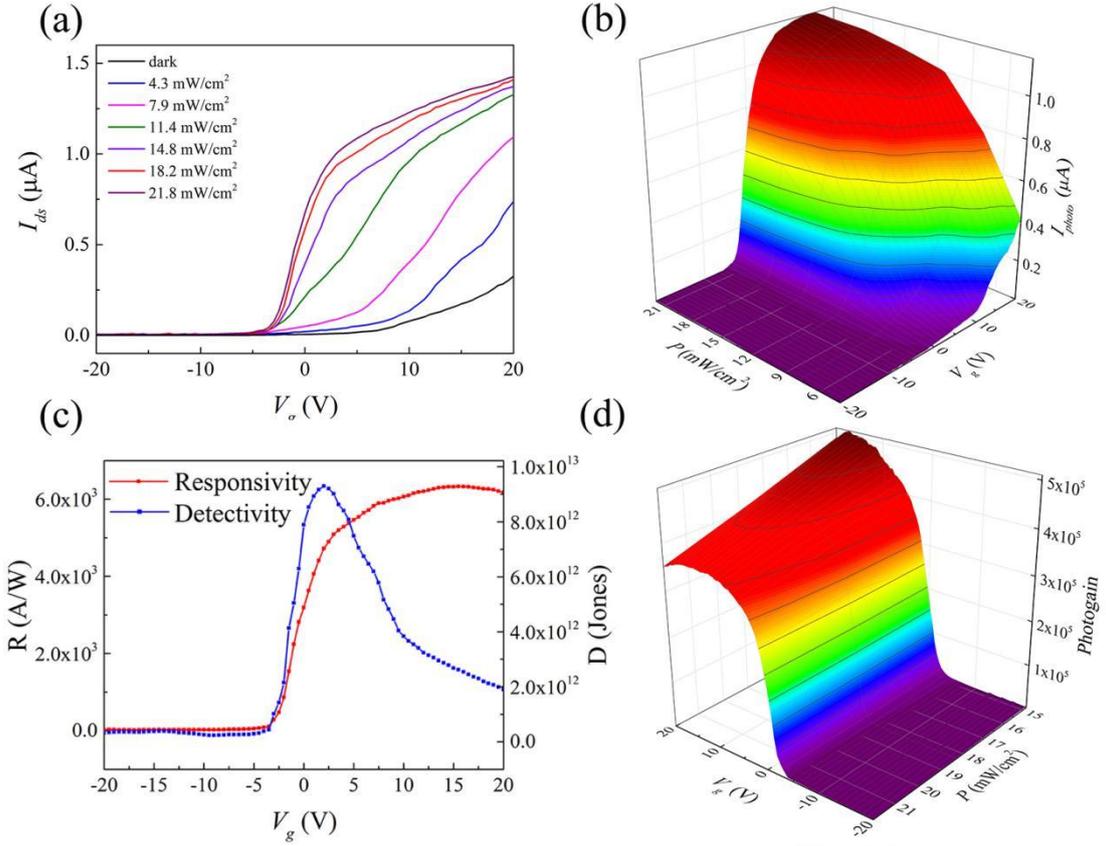

Figure 4. (a) Transfer curves of the device recorded in the dark and under different illumination intensities at $V_{ds}$ = 1 V. (b) 3D view of photocurrent mapping. (c) The responsivity and detectivity as a function of $V_g$ measured at $V_{ds}$ = 1 V. (d) 3D view of photogain mapping at $V_{ds}$ = 1 V.